\documentclass[apjl]{emulateapj}
\usepackage{epsfig}
\begin{document}

\title{Is the Milky Way ringing? The hunt for high velocity streams}

\author{I.~Minchev\altaffilmark{1}, A.~C.~Quillen\altaffilmark{2},
M.~Williams\altaffilmark{3,4}, K.~C.Freeman\altaffilmark{5}, J.~Nordhaus\altaffilmark{6}, 
A.~Siebert\altaffilmark{1}, and O.~Bienaym\'{e}\altaffilmark{1}
}

\altaffiltext{1}{Universit\'e de Strasbourg, CNRS, Observatoire Astronomique, 
Strasbourg, France; minchev@astro.u-strasbg.fr}
\altaffiltext{2}{Department of Physics and Astronomy, University of Rochester,
Rochester, NY 14627}
\altaffiltext{3}{Mount Stromlo Observatory, Australian National University,
Cotter Road, ACT 2611, Australia}
\altaffiltext{4}{Astrophysikalisches Institut Potsdam, An der Sternwarte 16,
D-14482, Potsdam, Germany} 
\altaffiltext{5}{Australian National University, Canberra, Australia}
\altaffiltext{6}{Department of Astrophysical Sciences, Princeton University,
Princeton, NJ 08544}

\begin{abstract}
We perform numerical simulations of a stellar galactic disk with
initial conditions chosen to represent an unrelaxed population which 
might have been left following a merger. Stars are unevenly distributed 
in radial action angle, though the disk is axisymmetric. The velocity 
distribution in the simulated Solar neighborhood exhibits waves traveling 
in the direction of positive $v$, where $u,v$ are the radial and tangential  
velocity components. As the system relaxes and structure wraps in phase space, 
the features seen in the $uv$-plane move closer together. We show that these
results can be obtained also by a semi-analytical method. We propose
that this model could provide an explanation for the high velocity streams
seen in the Solar neighborhood at approximate $v$ in km/s, of -60 (HR 1614),
-80 \citep{arifyanto06}, -100 (Arcturus), and -160 \citep{klement08}. In
addition, we predict four new features at $v\approx$ -140, -120, 40 and 60 km/s.
By matching the number and positions of the observed streams, we estimate that 
the Milky Way disk was strongly perturbed $\sim$1.9 Gyr ago. This event 
could have been associated with Galactic bar formation.
\end{abstract}


\section{Introduction}

The formation and evolution of galaxies is one of the most important
topics in contemporary astrophysics. High-redshift cosmology provides insight
into the evolution of global galaxy properties, but is unable to probe the
internal kinematics and chemistry on sub-galactic scales. The Milky Way (MW), on 
the other hand, contains a vast amount of fossil evidence encoded in the motions
and chemical properties of its stars. The Galaxy is the only galaxy within which 
we can obtain information at the level of detail required to distinguish robustly
between different formation scenarios. Unfortunately, however, the precise
structure of the MW remains a topic of debate. In order to make progress
in this field we need to differentiate between different Galactic models.

As part of this effort, models aimed at explaining asymmetries in the
Solar Neighborhood (SN) velocity space
as the result of internal (spiral and/or bar structure) or external (satellite
mergers) agents have been explored in the past several decades.
Hipparcos data revealed a non-smooth local velocity distribution of stars
\citep{dehnen98,chereul98,chereul99}. These stellar streams cannot simply be
dissolved clusters \citep{famaey07}, but could be caused by dynamical effects
within the MW disk or satellite mergers. 
Some of these features have been used as tracers of non-axisymmetric
Galactic disk structure and employed in estimating parameters of the 
MW central bar \citep{dehnen00,fux01,mnq07,chakrabarty07} and spirals 
\citep{lepine01,qm05,chakrabarty07}. However, dynamical instabilities in the 
Galactic disk have been found to only relate to velocities in the range 
$u,v\sim\pm50$ km/s, where $u,v$ are the radial and tangential galactocentric 
velocities of SN stars, respectively. At this time there is 
no evidence that spiral or bar structure can cause high velocity streams, thus, 
these are usually attributed to merger events. For instance, the Arcturus
stream at $v=-100$ km/s, has been interpreted as originating from the debris
of a disrupted satellite \citep{navaro04,helmi06}. Two recently discovered
streams at $v\sim80$ km/s \citep{arifyanto06} and $v\sim-160$ km/s
\citep{klement08} were assigned similar origin, based on their kinematics.
There is plenty of evidence for past and ongoing accretion of small objects by
the MW, the most dramatic one being the highly disrupted Sgr dwarf
galaxy identified by \citep{ibata94,ibata95}. 
But what about clues of more massive MW mergers? 

Cosmological simulations show that massive minor mergers are likely to have happened 
during the lifetime of a MW-sized host system. Early generations of numerical 
simulations \citep{quinn93,walker96} have investigated the heating of galactic 
disks by mergers, finding that disks heated in this way are similar to the MW 
thick disk. By considering realistic satellite orbits and
conditions, more recent attempts to model disk heating in a
cosmological context suggest smaller efficiency than previously thought.
For instance, \cite{hopkins08} estimated that the Milky Way could have survived 
as many as $\sim$5-10 minor (mass ratio to host disk $\sim$1:10) mergers in the 
last 10 Gyr. By estimating the maximum mass ratio merger for a MW-sized galaxy, 
the authors point out that the Galaxy could even have survived 
mergers of mass ratio $\sim1:4-1:3$ without destroying the disk.
\cite{kazantzidis08} estimated the number of massive subhaloes 
accreted since z$\sim 1$ to be at least one object with a mass $\sim M_{disc}$ 
and five objects more massive than 20\% $M_{disc}$. 
These findings are consistent with simulations by other groups 
\citep{benson04, stoehr06, delucia08,villalobos08,stewart08,purcell09}.
Even though massive minor mergers may be less frequent, they are able
to reach the center of the host system (provided they are also dense) 
thanks to dynamical friction,
causing important changes in the structure and kinematics of the
host disc. By using data from the Two Micron All-Sky Survey (2MASS),
\cite{cole02} estimated that the Milky Way bar is likely to have formed more 
recently than 3 Gyr ago and suggested that this event could have been triggered by 
a now-merged satellite. 
Formation of central bars as the result of satellite-disk encounters has also been
observed in N-body simulations \citep{walker96,kazantzidis08}.

All this evidence from both observations and simulations implies that 
merging satellites could contribute to the heating of the MW disc. Is it 
possible to relate the effects of such an 
event to features in the local velocity space ($uv$-plane)? Instead of 
interpreting streams as debris from disrupted small satellites, as done by 
\cite{navaro04,helmi06,arifyanto06,klement08}, we ask a different 
question here:
Is it possible that some overdensities in the SN velocity space are simply 
the response of the MW disk to the sudden energy kick imposed by a massive 
satellite in the past?

The signature of this perturbation will be present in the MW stellar
kinematics during the relaxation time after the event. Depending on the
time of this event, it is possible that the Galactic disk 
is still undergoing relaxation, although at this time 
the impact imprint may be extremely faint. 
However, because stars of the thick disk spend relatively little time near
the galactic plane, where the spiral arm heating and scattering by giant
molecular clouds is most vigorous, radial mixing within the thick disk is 
unlikely to wipe out the signature of a past event too quickly.

In this letter we look at the evolution of a non-relaxed galactic disk
and its manifestation in the velocity distribution of a simulated SN.

\begin{figure*}
\epsscale{1.2}
\plotone{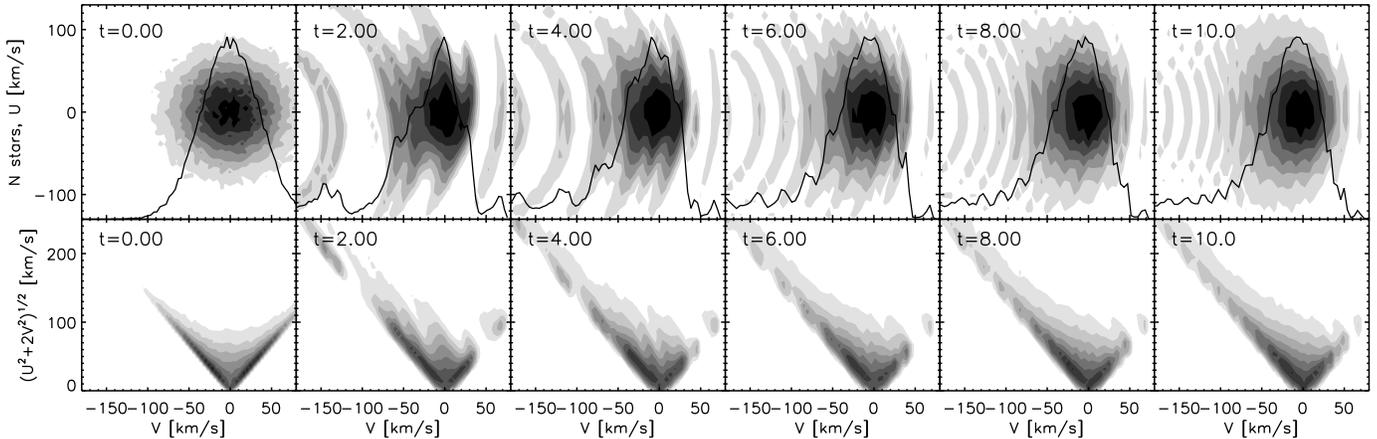}
\figcaption{
Time development of the local velocity field for an axisymmetric disk
with ICs as described in the text, presented in three different ways. 
First row shows the $uv$-plane (contours) and the tangential velocity distribution 
(solid line). The second row plots $(u^2+2v^2)^{1/2}$ versus $v$. 
The sample shown is limited to a radius of 100 pc from our 
fictitious Sun. We show six time outputs up to $t=10$ SN rotations. As time 
increases features get closer together as phase wrapping takes place. 
\label{fig:uv}
}
\end{figure*}
\begin{figure*}
\epsscale{1.2}
\plotone{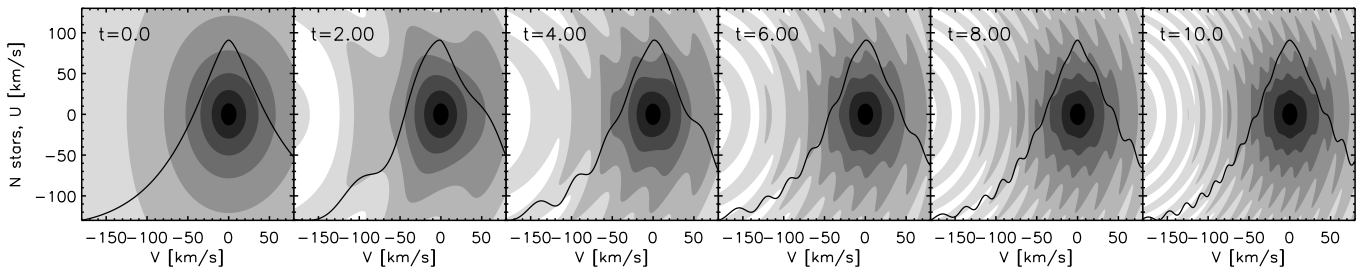}
\figcaption{
The $uv$-plane computed using a weighting function utilizing 
eq. 30b from \cite{dehnen99}. Note the striking similarity to our numerical
simulations (figure \ref{fig:uv}).
\label{fig:dehnen}
}
\end{figure*}

\section{Simulations of an unrelaxed disk}

At present it is not computationally feasible to achieve the statistics needed
for resolving a small spatial region with N-body simulations,
even in the case of a single galaxy. Thus, a realistic galaxy collision simulation
does not provide the resolution needed for our purposes, i.e., resolve a
fictitious SN to look for structure in the local velocity space.
We choose initial velocities for particles axisymmetrically by means of
Gaussian distributions in $u$ and $v$ with corresponding standard
deviations $\sigma_u$ and $\sigma_v$, consistent with a hot stellar population.
However, we purposely choose them so that they are not evenly distributed in 
their radial oscillation. This serves as a proxy for choosing a population
that is unrelaxed or unevenly distributed in phase space, such as might be left 
after a merger. We further simplify the problem by considering only two 
dimensions, assuming the vertical motion of stars is decoupled from the motion 
in the plane of the Galaxy. We are mainly concerned with a flat
rotation curve but also discuss the effect of a decreasing and increasing one
in section \ref{sec:concl}. In addition to an axisymmetric system we also 
simulate a disk perturbed by a central bar as a pure quadrupole. A detailed 
description of the perturbation can be found in \cite{mnq07}.  
To explore the time development of the system, we do not time-average over
position and velocity vectors, as it is frequently done in test-particle simulations
\citep{dehnen00,fux01,mq07} where no dynamical development of the system is 
expected. To convert to real units we use Local Standard of Rest (LSR) tangential 
velocity of 220 km/s, and Galactocentric distance of 7.8 kpc.

Note that the Gaussian distributions in $u$ and $v$ provide initial conditions
(ICs) sampled non-uniformly in $\theta$, the radial epicyclic angle. These ICs 
were found to induce an initial radial expansion in the disc consistent with the 
N-body simulations by \cite{quinn93} and \cite{walker96}, where they found that 
the host disks respond
by spreading both radially and vertically. However, after a couple of rotations 
the density distribution appears smooth and is axisymmetric.

\section{Results}
\label{sec:results}

\subsection{Density waves in velocity space}
\label{sec:waves}

We now look for the effect of our ICs on a simulated SN velocity distribution.
In figure \ref{fig:uv} we show the time development of the local velocity field
in three different ways. The first row shows the $uv$-plane (contours) and the 
$v$ distribution (solid line).
The second row plots $(u^2+2v^2)^{1/2}$ versus $v$ as done by \cite{arifyanto06}
and \cite{klement08}. In this simulation $\sigma_u=50$ km/s. The sample shown 
is limited to a radius of 100 pcs around 
our fictitious Sun. We show six time outputs up to $t=10$ rotations at $r_0$.  
Note that features in the $uv$-plane get closer together as time increases.
We interpret this as wrapping in phase space on a timescale associated with 
the epicyclic frequency. The features are not oriented along constant eccentricity 
surfaces as predicted by \cite{helmi06} for particles trapped from a minor merging
galaxy. In our case the arcs are oriented in the opposite direction, since they
are centered on $(u,v-v_0)=0$, in other words these are constant energy
surfaces.

Note that the features in the $uv$-plane are manifested in the tail of the 
$v$-distribution, as well as in $(u^2+2v^2)^{1/2}$, as overdensities traveling 
in the direction of positive $v$.  
In section \ref{sec:moving} these are used to match to high velocity streams
observed in the SN stellar population. 

It is important to make a point here concerning the average distance of our 
sample from the Sun. 
Because of the differential rotation of the Galaxy, at radii interior to $r_0$ 
the relaxation is completed faster, whereas the opposite is true for stellar
samples at $r>r_0$. Thus, for a given time, the separation of features
in the $uv$-plane depends on the Galactocentric distance of our sample.
Consequently, as sample depth increases we sample a large range of 
Galactic radii causing the waves to interfere and either enhance or (mostly) 
wipe out structure in the $uv$-plane.

To better understand the mechanism giving rise to these ripples in velocity
space, next we use the epicyclic approximation to reproduce the numerical 
results just described.  

\subsection{Semi-analytical approach}

We assume a flat rotation curve with the parameter
$\gamma \equiv 2 \Omega /\kappa  = \sqrt{2}$, where
$\Omega$ and $\kappa$ are the angular rotation rate
and epicyclic frequency of a particle in a circular orbit.
Since we work in units of $v_0$ and $r_0$, the gravitational potential for
a flat rotation curve $\Phi(r) = v_c^2 \ln(r/r_0)$ is zero for
stars in the SN. For every position on the $uv$-plane 
energy and angular momentum can be computed as
$E(u,v) = {1\over 2} (u^2 + (1+v)^2)$ and $L(v)=1+v$.

\citet{dehnen99} derived a good approximation for the frequency of epicycles
accurate at large epicyclic amplitude (his equation 30d):
\begin{equation}
\omega_R(E,L) \approx {\kappa(E) \over 
        \left[{1 + {1 \over 4}e^2(\gamma-1)(\gamma-2) }
        \right] },
\label{eqn:omegar}
\end{equation}
where
\begin{equation}
e^2 = 1 - L^2/L^2(E) 
\end{equation}
is the square of the orbital eccentricity (his equation 30b). These functions 
depend on $\kappa(E)$ and $L(E)$, which are the epicyclic frequency and the 
angular momentum for a circular orbit with energy $E$
\begin{eqnarray}
\kappa(E) &=& \sqrt{2} \exp \left({1\over 2} -E \right) \\
L(E)      &=& \exp{ \left( E - {1 \over 2} \right)},
\end{eqnarray}
where we have used expressions from Table 1 by \citet{dehnen99},
given in units of the circular velocity.
Since $\omega_R$ depends on $L$ and $E$, it can be computed
for every position on the $uv$-plane in the SN.

We assume that our initial particle distribution is not
evenly distributed in the angle associated with epicyclic motion, 
$\theta$. If the initial phase space density distribution is skewed 
along this angle at $\theta_0$, then at a time $\Delta t$ later there 
will be a maximum of particles at
\begin{equation}
\theta(L,E) = \theta_0 + \omega_R(L,E) \Delta t
\end{equation}
To mimic the effect of this we construct a weighting function
that gives a maximum at $\theta(L,E)=0~{\rm mod}~ 2 \pi$.
\begin{equation}
w(L,E) = {1\over 2} ( 1 + \cos(\omega_R t))
\end{equation}
Our initial phase space particle distribution is more unevenly distributed 
in action angle at large eccentricity so we can improve our weighting function
by multiplying the cosine function by the square of the orbital eccentricity.
\begin{equation}
w(L,E) = \exp(-e/e_0) ( 1 + e^2 \cos(\omega_R t)),
\label{eqn:weight}
\end{equation}
where the exponential function mimics a Gaussian velocity dispersion and $e_0$
is a constant. In the above equation $\omega_R$ and $e$ depend on $L,E$. 
Equation \ref{eqn:weight} can be computed for every position on the $uv$-plane 
for different values of $t$. The result is shown in figure \ref{fig:dehnen} 
and looks remarkably close to what is seen in the test particle simulations.

\subsection{Relating to moving groups}
\label{sec:moving}

\begin{figure}[t]
\epsscale{1.15}
\plotone{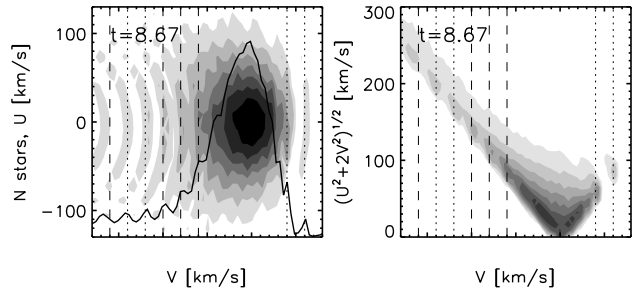}
\figcaption{
Three ways of plotting the same data, as done in figure \ref{fig:uv}, 
for a particular time matching four observed 
high velocity local streams (dashed lines). From left to right, in km/s, these are 
STR1 at $v\approx-160$, Arcturus at $v\approx-100$, STR2 at $v\approx-80$ and 
HR 1614 at $v\approx-60$. The dotted lines at $v\approx-140$, -120, 40 and 60 
km/s show the positions of four new stream this new model predicts.
Note that the density oscillations giving rise to these features have an amplitude 
of less than one percent.
\label{fig:match}
}
\end{figure}

We now discuss the possibility that ringing in the Galaxy is the reason for 
four high velocity streams observed in the local velocity field. We search 
for a particular time in our synthetic velocity distributions, for which we 
get a satisfactory match to all four. 
Figure \ref{fig:match} presents velocity field plots from our numerical
simulations for time $t=8.67$ 
rotations at $r_0$, in three different ways as in figure \ref{fig:uv}.
The dashed lines indicate the position of the observed overdensities. 
From left to right these are as follows:
(1) A high velocity stream at $v=-160\pm20$ km/s (hereafter STR1) was 
recently reported by \cite{klement08} which, based on its kinematics, 
is thought to belong to the thick disk. 
(2) Arcturus is a moving group lagging the LSR by 100 km/s. Its metal-poor
nature and significant age are consistent with the thick disk. A detailed
investigation of its origin by \cite{williams08} found that the chemical results
are consistent with a dynamical origin but do not entirely rule out a merger one.
The upper left-hand panel in figure 1 by \cite{williams08} presents a $uv$-plot
of the Arcturus stream. Centered on a narrow $v$, it spreads over the range
$-100<u<100$, consistent with our results (left panel in figure \ref{fig:match}). 
(3) A stream with characteristics appropriate for the thick disk at $v=-80$ km/s 
(hereafter STR2) was found by \cite{arifyanto06} using data extracted from 
various catalogues.  
(4) The moving group HR 1614 at $v$ = -60 km/s, is thought to be a dispersed 
open cluster because of its chemical homogeneity \citep{eggen92,desilva07} at a 
distance of 40 pc from the sun. It is intriguing that, similarly to Arcturus, 
in the $uv$-plane this stream has a spread in $u$, again consistent with figure 
\ref{fig:match}. However, $-50<u<20$ km/s (figure 5 in \citealt{desilva07}), 
i.e., the corresponding wave giving rise to HR 1614 is distorted toward negative 
$u$. This is consistent with the effect of the bar, given the proximity of this
overdensity to the Hercules stream. Note that our interpretation for HR 1614's 
location at $v=60$ km/s does not contradict the possibility of it being a 
dispersed cluster, as long as it is older than the time of the merger event,
since the streams in the model proposed here are only defined by their
kinematics.

In addition to the four observed streams, our new model predicts overdensities
at approximately every 20 km/s. However, in the range $-50<v<50$ km/s, the 
central peak of the velocity distribution dominates. 
Thus, we are left with four new, easily identifiable 
overdensities at $v\approx-140$, -120, 40 and 60 km/s, indicated by the dotted lines 
in figure \ref{fig:match}.

\begin{figure}[t]
\epsscale{1.2}
\plotone{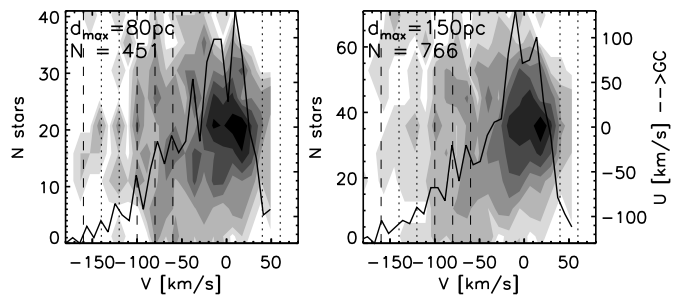}
\figcaption{
Observational data form \cite{nordstrom04} and \cite{schuster06} combined samples.
We select stars in the metallicity range $-1.1<$[Fe/H]$<-0.55$ dex. The dashed and
dotted lines indicate the observed and predicted overdensities as in figure
\ref{fig:match}.
\label{fig:obs}
}
\end{figure}

In order to look for the streams our model predicts, we combined the observational 
samples by \cite{nordstrom04} and \cite{schuster06}. We select a thick disk
dominated sample by considering the metallicity range $-1.1<$[Fe/H]$<-0.55$ dex.  
In figure \ref{fig:obs} we present the result for sample depths $d_{max}=80$ and 
$d_{max}=150$ pc in the same fashion as the left-hand panel of figure 
\ref{fig:match}. With this small number of stars (N=451,766 for 
$d_{max}=80,150$ pc), this result is not statistically significant to 
provide convincing evidence for the validity of our model. However, the 
resemblance of the observed $v$-distribution to the ones resulting from our 
simulations and semi-analytical approach, is striking.

\section{Discussion and conclusions}
\label{sec:concl}

We have shown that an axisymmetric galactic disk subjected to an initial energy 
kick approximating a massive Galactic merger induces waves in the SN velocity 
field propagating in the direction of positive $v$, which appear as 
overdensities in the tail of the tangential velocity distribution.
By comparing our synthetic models to observations, a satisfactory match to 
four SN high velocity streams is achieved toward the end 
of the disk relaxation at $t=8.67$ SN rotations (figure \ref{fig:match}). 
In addition, we predict the existence of four (or more) new features at 
$v\approx-140$, -120, 40 and 60 km/s.
Our results allow us to make an estimate for the time of the event.
For a Galactocentric distance of 7.8 kpc and a LSR tangential velocity of 220
km/s, $t=8.67$ rotations corresponds to $\sim1.9$ Gyr.
If our model is correct, then all observed and predicted streams must be older than 
the time of the disk stirring event. All four of the known features discussed 
in section \ref{sec:moving} have ages $>\sim2$ Gyr, which is consistent with
our prediction for the time of the impact.
This model also argues against purely diffusive stochastic heating models 
\citep{jenkins90,sellwood02,mq06} for the Galactic disk.

In addition to a flat rotation curve, we have also considered a power law initial
tangential velocity $v_\phi=v_0(r/r_0)^\beta$ with $\beta=0.2,-0.2$
corresponding to a rising and a declining rotation curve, respectively. 
We found that the separation of the features in the $uv$-plane decreases with
time more rapidly for $\beta=-0.2$ and more slowly in the case of $\beta=0.2$, 
as expected. However, at $r_0$ our results remained the same. The $\sim20$ 
km/s separation of features in the local velocity field arises naturally 
as a result of the Galactocentric distance and tangential velocity of the LSR, 
i.e., it is determined only by the LSR angular velocity. 
Thus, not only is this model independent of the MW rotation curve, but it can
be used to provide constraints on $\beta$ by observations of the velocity field at 
Galactic radii different than $r_0$. 

Features in the $uv$-plane represent curves of constant energy and are oriented 
opposite to the constant eccentricity curves in \cite{helmi06}'s model.
As more SN stars are surveyed with RAVE and GAIA, the shape of these will
be better resolved and we will be able to tell the difference between 
Helmi's model and ours. We predict a shift in location of features as a function
of Galactocentric radius (closer together for shorter radii)
as the distance between the features depends on the epicyclic frequency.
Deeper surveys, such as ARGOS, SEGUE, BRAVA, and APOGEE, could search for this 
shift in the way suggested by \cite{mq08}.

We have checked that the growth, or longer term effect, of a central bar does not 
cause similar features in the $uv$-plane, thus the bar is not responsible for 
such radial perturbations. However, an increase in central mass associated with a 
merger could cause such variations in the epicyclic action-angle distribution.
From recent cosmological simulations it is now known that satellites of mass
ratio to host disk $\geq 1:10$ merge on highly eccentric, nearly radial orbits in a 
couple of dynamical times \citep{hopkins08}. Then they quickly merge by dumping 
their mass in the center of the host disk.
It would appear to be simple to look for initial conditions from an N-body 
simulation. However this is actually nontrivial for a number of reasons:
(1) not enough particles to resolve high velocity structure;
(2) large range of possibilities;
(3) time dependent phenomena associated with mergers.

Our short timescale is consistent with the estimated age of the Galactic
bar, as measured by \cite{cole02} ($<3$ Gyr ago). This suggests that the same
event that caused the formation of the Galactic bar could have left the stellar 
disk unrelaxed, thus giving rise to the observed high velocity streams.
Another possibility for stirring up the MW disk is the $\omega$Cen event 
\citep{bekki03,meza05}.

Our choice of ICs for an unrelaxed disk is arbitrary.
N-body simulations could be explored to see if such
a distribution could arise from a merger and motivate better
choices of ICs for future particle integration studies.

\acknowledgements
We would like to thank Rodrigo Ibata and Dominique Aubert for helpful comments.
Support for this work was provided by ANR and RAVE.


\begin{thebibliography}{}

\bibitem[\protect\astroncite{{Arifyanto} and {Fuchs}}{2006}]{arifyanto06}
{Arifyanto}, M.~I. and {Fuchs}, B.: 2006,
\newblock {\em \aap} {\bf 449}, 533

\bibitem[\protect\astroncite{{Bekki} and {Freeman}}{2003}]{bekki03}
{Bekki}, K. and {Freeman}, K.~C.: 2003,
\newblock {\em \mnras} {\bf 346}, L11

\bibitem[\protect\astroncite{{Benson} et~al.}{2004}]{benson04}
{Benson}, A.~J., {Lacey}, C.~G., {Frenk}, C.~S., {Baugh}, C.~M., and {Cole},
  S.: 2004,
\newblock {\em \mnras} {\bf 351}, 1215

\bibitem[\protect\astroncite{{Chakrabarty}}{2007}]{chakrabarty07}
{Chakrabarty}, D.: 2007,
\newblock {\em \aap} {\bf 467}, 145

\bibitem[\protect\astroncite{{Chereul} et~al.}{1998}]{chereul98}
{Chereul}, E., {Creze}, M., and {Bienayme}, O.: 1998,
\newblock {\em \aap} {\bf 340}, 384

\bibitem[\protect\astroncite{{Chereul} et~al.}{1999}]{chereul99}
{Chereul}, E., {Cr{\'e}z{\'e}}, M., and {Bienaym{\'e}}, O.: 1999,
\newblock {\em \aaps} {\bf 135}, 5

\bibitem[\protect\astroncite{{Cole} and {Weinberg}}{2002}]{cole02}
{Cole}, A.~A. and {Weinberg}, M.~D.: 2002,
\newblock {\em \apjl} {\bf 574}, L43

\bibitem[\protect\astroncite{{De Lucia} and {Helmi}}{2008}]{delucia08}
{De Lucia}, G. and {Helmi}, A.: 2008,
\newblock {\em \mnras} {\bf 391}, 14

\bibitem[\protect\astroncite{{De Silva} et~al.}{2007}]{desilva07}
{De Silva}, G.~M., {Freeman}, K.~C., {Bland-Hawthorn}, J., {Asplund}, M., and
  {Bessell}, M.~S.: 2007,
\newblock {\em \aj} {\bf 133}, 694

\bibitem[\protect\astroncite{{Dehnen}}{1998}]{dehnen98}
{Dehnen}, W.: 1998,
\newblock {\em \aj} {\bf 115}, 2384

\bibitem[\protect\astroncite{{Dehnen}}{1999}]{dehnen99}
{Dehnen}, W.: 1999,
\newblock {\em \aj} {\bf 118}, 1190

\bibitem[\protect\astroncite{{Dehnen}}{2000}]{dehnen00}
{Dehnen}, W.: 2000,
\newblock {\em \aj} {\bf 119}, 800

\bibitem[\protect\astroncite{{Eggen}}{1992}]{eggen92}
{Eggen}, O.~J.: 1992,
\newblock {\em \aj} {\bf 104}, 1906

\bibitem[\protect\astroncite{{Famaey} et~al.}{2007}]{famaey07}
{Famaey}, B., {Pont}, F., {Luri}, X., {Udry}, S., {Mayor}, M., and {Jorissen},
  A.: 2007,
\newblock {\em \aap} {\bf 461}, 957

\bibitem[\protect\astroncite{{Fux}}{2001}]{fux01}
{Fux}, R.: 2001,
\newblock {\em \aap} {\bf 373}, 511

\bibitem[\protect\astroncite{{Helmi} et~al.}{2006}]{helmi06}
{Helmi}, A., {Navarro}, J.~F., {Nordstr{\"o}m}, B., {Holmberg}, J., {Abadi},
  M.~G., and {Steinmetz}, M.: 2006,
\newblock {\em \mnras} {\bf 365}, 1309

\bibitem[\protect\astroncite{{Hopkins} et~al.}{2008}]{hopkins08}
{Hopkins}, P.~F., {Hernquist}, L., {Cox}, T.~J., {Younger}, J.~D., and {Besla},
  G.: 2008,
\newblock {\em \apj} {\bf 688}, 757

\bibitem[\protect\astroncite{{Ibata} et~al.}{1994}]{ibata94}
{Ibata}, R.~A., {Gilmore}, G., and {Irwin}, M.~J.: 1994,
\newblock {\em \nat} {\bf 370}, 194

\bibitem[\protect\astroncite{{Ibata} et~al.}{1995}]{ibata95}
{Ibata}, R.~A., {Gilmore}, G., and {Irwin}, M.~J.: 1995,
\newblock {\em \mnras} {\bf 277}, 781

\bibitem[\protect\astroncite{{Jenkins} and {Binney}}{1990}]{jenkins90}
{Jenkins}, A. and {Binney}, J.: 1990,
\newblock {\em \mnras} {\bf 245}, 305

\bibitem[\protect\astroncite{{Kazantzidis} et~al.}{2008}]{kazantzidis08}
{Kazantzidis}, S., {Bullock}, J.~S., {Zentner}, A.~R., {Kravtsov}, A.~V., and
  {Moustakas}, L.~A.: 2008,
\newblock {\em \apj} {\bf 688}, 254

\bibitem[\protect\astroncite{{Klement} et~al.}{2008}]{klement08}
{Klement}, R., {Fuchs}, B., and {Rix}, H.-W.: 2008,
\newblock {\em \apj} {\bf 685}, 261

\bibitem[\protect\astroncite{{L{\'e}pine} et~al.}{2001}]{lepine01}
{L{\'e}pine}, J.~R.~D., {Mishurov}, Y.~N., and {Dedikov}, S.~Y.: 2001,
\newblock {\em \apj} {\bf 546}, 234

\bibitem[\protect\astroncite{{Meza} et~al.}{2005}]{meza05}
{Meza}, A., {Navarro}, J.~F., {Abadi}, M.~G., and {Steinmetz}, M.: 2005,
\newblock {\em \mnras} {\bf 359}, 93

\bibitem[\protect\astroncite{{Minchev} et~al.}{2007}]{mnq07}
{Minchev}, I., {Nordhaus}, J., and {Quillen}, A.~C.: 2007,
\newblock {\em \apjl} {\bf 664}, L31

\bibitem[\protect\astroncite{{Minchev} and {Quillen}}{2006}]{mq06}
{Minchev}, I. and {Quillen}, A.~C.: 2006,
\newblock {\em \mnras} {\bf 368}, 623

\bibitem[\protect\astroncite{{Minchev} and {Quillen}}{2007}]{mq07}
{Minchev}, I. and {Quillen}, A.~C.: 2007,
\newblock {\em \mnras} {\bf 377}, 1163

\bibitem[\protect\astroncite{{Minchev} and {Quillen}}{2008}]{mq08}
{Minchev}, I. and {Quillen}, A.~C.: 2008,
\newblock {\em \mnras} {\bf 386}, 1579

\bibitem[\protect\astroncite{{Navarro} et~al.}{2004}]{navaro04}
{Navarro}, J.~F., {Helmi}, A., and {Freeman}, K.~C.: 2004,
\newblock {\em \apjl} {\bf 601}, L43

\bibitem[\protect\astroncite{{Nordstr{\"o}m} et~al.}{2004}]{nordstrom04}
{Nordstr{\"o}m}, B., {Mayor}, M., {Andersen}, J., {Holmberg}, J., {Pont}, F.,
  {J{\o}rgensen}, B.~R., {Olsen}, E.~H., {Udry}, S., and {Mowlavi}, N.: 2004,
\newblock {\em \aap} {\bf 418}, 989

\bibitem[\protect\astroncite{{Purcell} et~al.}{2009}]{purcell09}
{Purcell}, C.~W., {Kazantzidis}, S., and {Bullock}, J.~S.: 2009,
\newblock {\em \apjl} {\bf 694}, L98

\bibitem[\protect\astroncite{{Quillen} and {Minchev}}{2005}]{qm05}
{Quillen}, A.~C. and {Minchev}, I.: 2005,
\newblock {\em \aj} {\bf 130}, 576

\bibitem[\protect\astroncite{{Quinn} et~al.}{1993}]{quinn93}
{Quinn}, P.~J., {Hernquist}, L., and {Fullagar}, D.~P.: 1993,
\newblock {\em \apj} {\bf 403}, 74

\bibitem[\protect\astroncite{{Schuster} et~al.}{2006}]{schuster06}
{Schuster}, W.~J., {Moitinho}, A., {M{\'a}rquez}, A., {Parrao}, L., and
  {Covarrubias}, E.: 2006,
\newblock {\em \aap} {\bf 445}, 939

\bibitem[\protect\astroncite{{Sellwood} and {Binney}}{2002}]{sellwood02}
{Sellwood}, J.~A. and {Binney}, J.~J.: 2002,
\newblock {\em \mnras} {\bf 336}, 785

\bibitem[\protect\astroncite{{Stewart} et~al.}{2008}]{stewart08}
{Stewart}, K.~R., {Bullock}, J.~S., {Wechsler}, R.~H., {Maller}, A.~H., and
  {Zentner}, A.~R.: 2008,
\newblock {\em \apj} {\bf 683}, 597

\bibitem[\protect\astroncite{{Stoehr}}{2006}]{stoehr06}
{Stoehr}, F.: 2006,
\newblock {\em \mnras} {\bf 365}, 147

\bibitem[\protect\astroncite{{Villalobos} and {Helmi}}{2008}]{villalobos08}
{Villalobos}, {\'A}. and {Helmi}, A.: 2008,
\newblock {\em \mnras} pp 1332--+

\bibitem[\protect\astroncite{{Walker} et~al.}{1996}]{walker96}
{Walker}, I.~R., {Mihos}, J.~C., and {Hernquist}, L.: 1996,
\newblock {\em \apj} {\bf 460}, 121

\bibitem[\protect\astroncite{{Williams} et~al.}{2008}]{williams08}
{Williams}, M.~E.~K., {Freeman}, K.~C., {Helmi}, A., and {the RAVE
  collaboration}: 2008,
\newblock {\em ArXiv e-prints}

\end{thebibliography}
\end{document}